\documentclass[11pt]{article}%
\usepackage{amsmath}
\usepackage{amsthm}
\usepackage{dsfont}
\usepackage{graphicx}
\usepackage{mathdots}
\usepackage{xcolor}%
\usepackage{amsfonts}%
\usepackage{amssymb}
\newcommand{\eqnb}{\begin{equation}}
\newcommand{\eqne}{\end{equation}}

\newtheorem{The}{Theorem}

\newtheorem{Rem}{Remark}

\begin{document}

\title{Stochastic Performance Modeling for Practical Byzantine Fault Tolerance
Consensus in Blockchain}
\author{Fan-Qi Ma$^{a}$, Quan-Lin Li$^{b}$ \thanks{Corresponding author: Q.L. Li
(liquanlin@tsinghua.edu.cn)}, Yi-Han Liu$^{b}$, Yan-Xia Chang$^{b}$\\$^{a}$School of Economics and Management,\\Yanshan University, Qinhuangdao 066004, China\\$^{b}$School of Economics and Management,\\Beijing University of Technology, Beijing 100124, China}
\maketitle

\begin{abstract}
The practical Byzantine fault tolerant (PBFT) consensus mechanism is one of
the most basic consensus algorithms (or protocols) in blockchain technologies,
thus its performance evaluation is an interesting and challenging topic due to
a higher complexity of its consensus work in the peer-to-peer network. This
paper describes a simple stochastic performance model of the PBFT consensus
mechanism, which is refined as not only a queueing system with complicated
service times but also a level-independent quasi-birth-and-death (QBD)
process. From the level-independent QBD process, we apply the matrix-geometric
solution to obtain a necessary and sufficient condition under which the PBFT
consensus system is stable, and to be able to numerically compute the
stationary probability vector of the QBD process. Thus we provide four useful
performance measures of the PBFT consensus mechanism, and can numerically
calculate the four performance measures. Finally, we use some numerical
examples to verify the validity of our theoretical results, and show how the
four performance measures are influenced by some key parameters of the PBFT
consensus. By means of the theory of multi-dimensional Markov processes, we
are optimistic that the methodology and results given in this paper are
applicable in a wide range research of PBFT consensus mechanism and even other
types of consensus mechanisms.

\vskip                    0.5cm


\end{abstract}

\section{Introduction}

The past decade has witnessed rapid development and growing popularity of
blockchain technologies. This has been attracting tremendous interests and
enthusiasm from both research communities and industrial applications. The
blockchain technologies were originated from a digital financial sector as a
decentralized, immutable, auditable, accountability ledger system in order to
deal with daily transactional data. So far it has been envisioned as a
powerful backbone/framework for decentralized data processing and data-driven
autonomous organization in a peer-to-peer and open-access network. For
blockchain technologies, readers may refer to books by Narayanan et al.
\cite{Nar:2016}, Bashir \cite{Bas:2018}, Raj \cite{Raj:2019}, Maleh et al.
\cite{Mal:2020}, Rehan and Rehmani \cite{Reh:2020} and Schar and Berentsen
\cite{Sch:2020}; and survey papers by Fauziah et al. \cite{Fau:2020} for smart
contracts, Sharma et al. \cite{Sha:2020} for cloud computing, Ekramifard et
al. \cite{Ekr:2020} for AI, Dai et al. \cite{Dai:2020} for IoT and Huang et
al. \cite{Hua:2021}.

The consensus mechanisms are always an important direction in the research of
blockchain technologies. Up to now, there have been more than 50 different
consensus mechanisms developed in blockchain technologies. We refer readers to
recent survey papers by, for example, Cachin and Vukoli\'{c} \cite{Cac:2017},
Bano et al. \cite{Ban:2017}, Natoli et al. \cite{Nat:2017}, Chaudhry and
Yousaf \cite{Cha:2018}, Nguyen and Kim \cite{Ngu:2018}, Salimitari and
Chatterjee \cite{Sal:2018}, Wang et al. \cite{Wan:2019}, Pahlajani et al.
\cite{Pah:2019}, Carrara et al. \cite{Car:2020}, Wan et al. \cite{Wan:2020},
Xiao et al. \cite{Xia:2020}, Ferdous et al. \cite{Fer:2020}, Nijsse and
Litchfield \cite{Nij:2020}, Yao et al. \cite{Yao:2021} and Khamar and Patel
\cite{Kha:2021}.

The PBFT consensus mechanism is the most basic one of blockchain consensus
mechanisms, and it plays a key role in extending, generalizing and developing
new effective blockchain consensus mechanisms. A reliable computer system must
be able to cope with the failure of one or more of its components, and a
failure components can send conflicting information to different parts of the
computer system. In this case, solving the type of failure and conflicting
problems is called a \textit{Byzantine generals problem}. See Lamport et al.
\cite{Lam:1982}, Lamport \cite{Lam:1983} and Martin and Alvisi \cite{Mar:2016}%
. Based on the Byzantine generals problem, Pease et al. \cite{Pease:1980} and
Lamport \cite{Lam:1982} provided the Byzantine fault tolerant (BFT) consensus
mechanism. To prevent malicious attacks and guarantee security of blockchain,
Castro and Liskov \cite{Castro:1999} proposed the PBFT consensus mechanism.
Thereafter, some researchers further developed various different PBFT
consensus mechanisms to effectively improve the performance of the PBFT
consensus mechanism. Important examples include Castro and Liskov
\cite{Cas:2002}, Veronese et al. \cite{Veronese:2011}, Abraham et al.
\cite{Abr:2017}, Kiayias and Russell \cite{Kiayias:2018}, Hao et al.
\cite{Hao:2018}, Gueta et al. \cite{Gueta:2019}, Malkhi et al.
\cite{Malkhi:2019}, Bravo et al. \cite{Bra:2020}, Sakho et al. \cite{Sak:2020}%
, Meshcheryakov et al. \cite{Mes:2021}, and Alqahtani and Demirbas
\cite{Alq:2021}.

It is an interesting topic to provide stochastic models for performance
evaluation of PBFT consensus mechanism. To our best knowledge, this paper is
the first one to set up such a stochastic model for performance evaluation of
PBFT consensus mechanisms. To this end, we first set up a queueing model with
more complicated service times, which are due to a higher complexity of block
validation. Then we express the PBFT queue as a level-independent QBD process,
which is a two-dimensional Markov process. By using the matrix-geometric
solution, we can provide a more detailed analysis for performance evaluation
of PBFT consensus mechanism, in which four performance measures are provided.
It is worthwhile to note that Li et al. \cite{Li:2018, Li:2019} are two
closely related works which gave performance analysis of the Proof of Work
(another more important consensus mechanism). Obviously, the PBFT consensus
mechanism makes the queueing model, together with the level-independent QBD
process, more complicated than that of the Proof of Work. Therefore, this
paper, together with Li et al. \cite{Li:2018, Li:2019}, can be regarded as
some key research on the matrix-analytic method in performance evaluation of
blockchain consensus mechanisms. Also, we believe that the matrix-analytic
method can play a key role in finding a more complete solution of stochastic
models in the study of blockchain technologies.

It is clear that the Markov processes and queueing theory play a key role in
the study of blockchain technologies. Readers may refer to survey papers by,
for example, Li et al. \cite{Li:2019}, Smetanin et al. \cite{Sme:2020}, Fan et
al. \cite{Fan:2020} and Huang et al. \cite{Hua:2021}. Based on this, it is
necessary and useful to review some available literature in this area as follows:

\textbf{(a) }\textit{Markov processes of blockchain systems: }Up to now, few
papers have applied the Markov processes (or Markov chains) to analysis of
blockchain systems. See Li et al. \cite{Li:2019} for some early research. For
some research well related to this paper, G\"{o}bel et al. \cite{Goebel:2016}
used a two-dimensional Markov process to provide a computational method for
dealing with the influence of propagation delay on the blockchain evolution.
Also, a new computational method was further developed by Javier and Fralix
\cite{Jav:2020}. Li et al. \cite{Li:2020} provided a new theoretical framework
of pyramid markov processes for blockchain selfish mining, where a new
matrix-geometric solution is developed and is different from that used in Li
et al. \cite{Li:2018, Li:2019}.

Nayak et al. \cite{Nayak:2016} constructed three different Markov processes to
analyze the stubborn mining strategies and designed a provable security
consensus protocol. Kiffer et al. \cite{Kiffer:2018} proposed a simple
consistency method of blockchain protocols by using a Markov chain. Huang et
al. \cite{Huang:2019} applied the Markov processes to performance evaluation
of the consistency algorithm Raft in the blockchain network. Carlsten
\cite{Carlsten:2016} used a Markov process to analyze the influence of
transaction fees on the blockchain selfish mining. Bai et al. \cite{Bai:2019}
used the Markov chains to discuss how the existence of multiple misbehaving
pools influences the profitability of selfish mining. Li et al.
\cite{LiY:2020} applied the Markov processes to provide performance and
security analysis for the direct acyclic graph-based ledger for Internet of
Things. Li et al. \cite{Li:2021} employed the Markov processes to the block
access control in wireless blockchain network.

\textbf{(b) }\textit{Queueing models of blockchain systems: }Li et al.
\cite{Li:2019} provided an overview for queueing models of blockchain
systems\textit{.} From that time forward, some new literatures are further
listed, such as Gopalan et al. \cite{Gop:2020}, Papadis et al. \cite{Pap:2018}%
, Geissler \cite{Gei:2019}, Mi\v{s}i\'{c} et al. \cite{Mis:2020}, Fang and Liu
\cite{FanM:2020}, Fralix \cite{Fra:2020}, Varma and Maguluri \cite{Var:2021}
and Wilhelmi and Giupponi \cite{Wil:2021}.

Based on the above analysis, the main contributions of this paper are
summarized as:

\begin{enumerate}
\item This paper describes a simple stochastic performance model of the PBFT
consensus mechanism, which is refined as not only a queueing system with
complicated service times but also a level-independent QBD process.

\item By applying the matrix-geometric solution, this paper obtains a
necessary and sufficient condition under which the PBFT consensus system is
stable, and provides the stationary probability vector of the QBD process,
which is used to set up and numerically compute four useful performance
measures of the PBFT consensus mechanism.

\item Some numerical examples are used to verify the validity of our
theoretical results, and to show how the four performance measures are
influenced by some key parameters of the PBFT consensus mechanism. Although
our Markov queueing model is simply designed for the PBFT consensus mechanism,
our analytic method will open a series of potentially promising research in
queueing theory and Markov processes of PBFT consensus mechanisms.
\end{enumerate}

The rest of this paper is organized as follows. Section 2 describes a
stochastic performance model of PBFT consensus mechanism.\ Section 3 expresses
the stochastic performance model as a level-independent QBD process. By using
the matrix-geometric solution, we obtain a necessary sufficient condition and
the stationary probability vector of the QBD process, and provide four
performance measures of the PBFT consensus mechanism. Section 4 uses some
numerical examples to verify the validity of our theoretical results, and
shows how the four performance measures are influenced by key system
parameters. Some concluding remarks are given in Section 5.

\section{A Stochatic Performance Model}

In this section, we describe a simple stochastic performance model of PBFT
consensus mechanism based on the operations structure of PBFT consensus
mechanism, in which several key factors are assumed to be exponential
distributions or Poisson processes. Also, we introduce some necessary
notations used in our later study.

In the PBFT consensus mechanism, there are $N$ nodes in the blockchain
network, and the number of Byzantine nodes is no more than $f$. Let $N=3f+1$.
To set up a useful relationship between the PBFT consensus mechanism and a
queueing model, we assume that there are always a lot of transactions in the
transaction pool, thus a number of transactions is packed into a file package.
These transaction packages are made one by one, and they are regarded as an
arrival process at the client. See Figure 1 for more details.

\begin{figure}[tbh]
\centering    \includegraphics[height=12cm
,width=15cm]{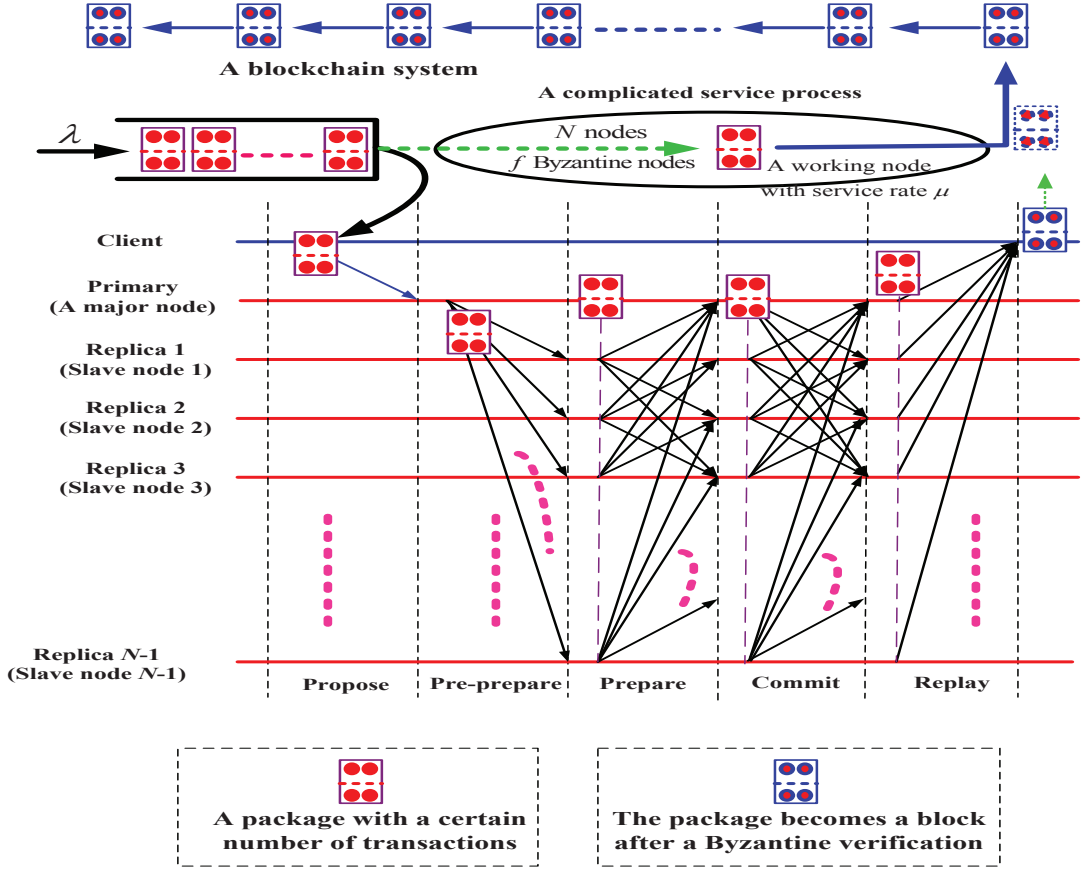}
\caption{A useful relationship between the PBFT consensus and a queueing
model}%
\end{figure}

Based on Figure 1, we provide a detailed description for the stochastic
performance model of PBFT consensus mechanism as follows:

\textbf{(1) The package arrival process}

We assume that the arrivals of many transaction packages at the client is a
Poisson process with arrival rate $\lambda$ for $\lambda>0$.

\textbf{(2) A major node and }$N-1$\textbf{ slave nodes}

A major node can be chosen from the $N$ nodes with equal probability. Once a
node is the major node, all the other nodes become $N-1$ slave nodes. On the
other hand, once a block is pegged on the blockchain by using the PBFT
consensus mechanism, the major node can receive a reward $c$ from the
blockchain system.

\textbf{(3) The block-generated and block-pegged process}

Once a transaction package arrives at the client, it is immediately submitted
to a chosen major node, and the major node also immediately submits the
transaction package to each of $N-1$ slave nodes. In this situation, the major
node and the $N-1$ slave nodes begin to deal with the transaction package
through the following three stages of PBFT consensus mechanism:
\textit{Prepare}, \textit{commit} and \textit{reply}, e.g., see Figure 1 for
their details. Now, we further explain these three stages as follows:

\textit{Prepare}:\textbf{ }Once one of the $N-1$ slaves receives the message
of the transaction package from the major node, it verifies the message
content to ensure that the message content has not been tampered with during
transmission. After the message is validated correctly, the slave node
immediately sends a preparation message to all other nodes except itself, that
is, $N-2$ slave nodes and the major node. When $2f+1$ different nodes receive
that the message is consistent with that of the major node, the next
\textit{commit} stage will be started immediately. In fact, the major node has
the same work as that of the $N-2$ slave nodes in this stage.

\textit{Commit}:\textbf{ }Once the \textit{prepare} stage is finished, each
node sends a \textit{commit }message to other nodes except itself. After other
nodes receive the \textit{commit} message, they verify the message content.
When the node has received $2f+1$ \textit{commit} messages including itself
and the message content is consistent with that of the previous
\textit{prepare} stage, the \textit{commit} message is verified and enters the
next \textit{reply} stage.

\textit{Reply}:\textbf{ }Each node sends a \textit{reply }message to the
client. If the client receives the same \textit{reply }messages sent by $2f+1$
different nodes, then it is regarded as that all the nodes (the major node and
the $N-1$ slave nodes) have reached a round of consensus. Therefore, the
transaction package becomes a legitimate block which can be pegged on the blockchain.

For simplicity of analysis, we assume that for each of the $N$ nodes (the
major node and the $N-1$ slave nodes), the time duration of going through
these three stages (\textit{prepare}, \textit{commit} and\ \textit{reply}) is
exponential with mean $1/\mu$ for $\mu>0$.

\textbf{(4) Independence}

We assume that all the random variables defined above are independent of each other.

\begin{Rem}
If the three stages (\textit{prepare}, \textit{commit} and\ \textit{reply})
are assumed to be exponential with means $1/\mu_{1}$, $1/\mu_{2}$ and
$1/\mu_{3}$, respectively, then the total time duration of going through the
three stages is a generalized Erlang distribution of order 3. In this case,
analysis of the stochastic performance model will be more complicated and is
regarded as one of our future studies. Therefore, this paper uses a simple
model to express a clear operations structure and four useful performance
measures in the PBFT consensus mechanism.
\end{Rem}

\section{A QBD process}

In this section, we express the stochastic performance model as a
level-independent QBD process, which is a two-dimensional continuous-time
Markov process. By using the matrix-geometric solution, we obtain a necessary
sufficient stable condition of the QBD process and its stationary probability
vector. Based on this, we provide four performance measures of the PBFT
consensus mechanism.

In the PBFT consensus mechanism, let $K(t)$ be the number of transaction
packages which are waiting at the client and will be submitted to the major
node one by one at time $t$. Let $M(t)$ be the number of nodes who have
verified the validity of one transaction package in the PBFT consensus
mechanism at time $t$. It is easy to see that $K(t)\in\left\{  0,1,2,\ldots
\right\}  $ and $M(t)\in\left\{  0,1,2,\ldots,2f\right\}  $. Then $\left\{
\left(  K(t),M(t)\right)  :t\geq0\right\}  $ is a two-dimensional
continuous-time Markov process whose state transition relations are shown in
Figure $2$.

\begin{figure}[th]
\centering      \includegraphics[width=10cm]{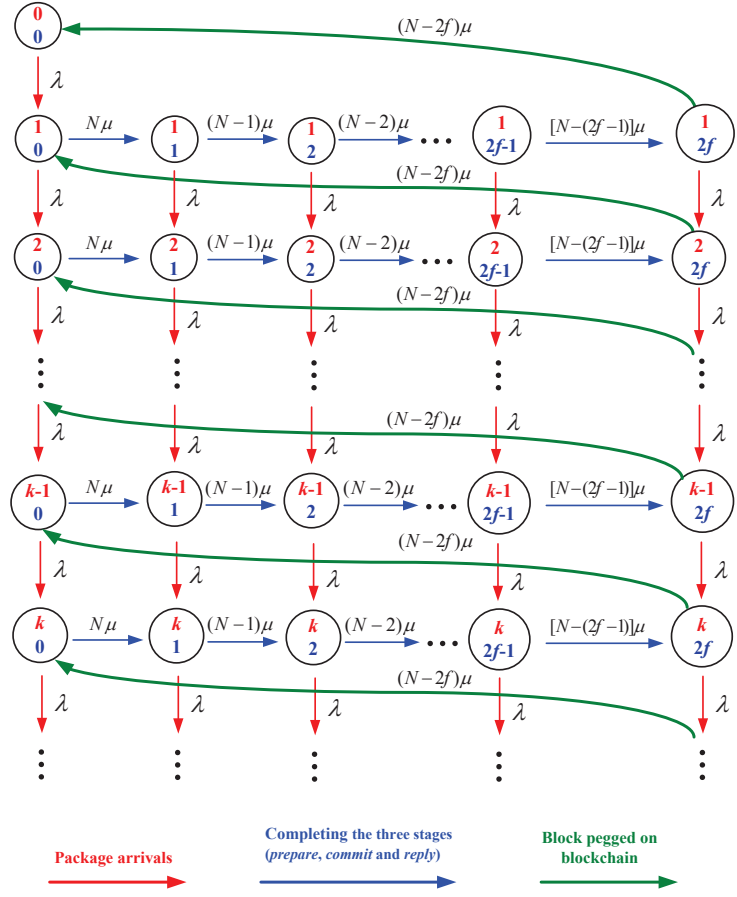}  \newline
\caption{The state transition relations of the Markov process}%
\label{figure:figure-2}%
\end{figure}

From Figure $2$, it is easy to see that the state space of the Markov process
$\left\{  \left(  K(t),M(t)\right)  :t\geq0\right\}  $ is given by%

\begin{align*}
\Omega &  =\left\{  (k\text{ (red)},m\text{ (blue)}):k=0,1,2,\ldots;\text{
}m=0,1,2,\ldots,2f\right\} \\
&  =\bigcup\limits_{k=0}^{\infty}\text{Level }k
\end{align*}
where%
\[
\text{Level }k=\left\{  (k,0),(k,1),(k,2),\ldots,(k,2f)\right\}  .
\]

Based on the state space $\Omega=\cup_{k=0}^{\infty}$Level $k$, the
infinitesimal generator of the Markov process $\left\{  \left(
K(t),M(t)\right)  :t\geq0\right\}  $ is written as%

\begin{equation}
Q=\left(
\begin{array}
[c]{ccccc}%
B_{1} & B_{0} &  &  & \\
B_{2} & A_{1} & A_{0} &  & \\
& A_{2} & A_{1} & A_{0} & \\
&  & \ddots & \ddots & \ddots
\end{array}
\right)  , \label{Gener}%
\end{equation}
where%

\[
B_{1}=-\lambda,B_{0}=\lambda(1,0,0,\ldots,0),
\]%

\[
B_{2}=(N-2f)\mu(0,0,\ldots,0,1)^{T},
\]%
\[
A{\small {_{1}=}}\left(
\begin{array}
[c]{ccccc}%
-N\mu-\lambda & N\mu &  &  & \\
& -\left(  N-1\right)  \mu-\lambda & \left(  N-1\right)  \mu &  & \\
&  & \ddots & \ddots & \\
&  &  & -(N-2f+1)\mu-\lambda & (N-2f+1)\mu\\
&  &  &  & -(N-2f)\mu-\lambda
\end{array}
\right)
\]%

\[
A_{0}=\left(
\begin{array}
[c]{ccccc}%
\lambda &  &  &  & \\
& \lambda &  &  & \\
&  & \ddots &  & \\
&  &  & \lambda & \\
&  &  &  & \lambda
\end{array}
\right)  ,\text{ \ }A_{2}=\left(
\begin{array}
[c]{ccccc}
&  &  &  & \\
&  &  &  & \\
&  &  &  & \\
&  &  &  & \\
(N-2f)\mu &  &  &  &
\end{array}
\right)  .
\]

The following theorem provides a necessary and sufficient condition under
which the QBD process $Q$ is stable. Based on this, we can obtain the system
stability of the PBFT consensus mechanism.

\begin{The}
The QBD process $Q$ is stable if and only if $\rho=\left(  \lambda/\mu\right)
\sum_{k=0}^{2f}1/\left(  N-k\right)  <1$. Thus, the stochastic system of the
PBFT consensus mechanism is stable if and only if $\rho=\left(  \lambda
/\mu\right)  \sum_{k=0}^{2f}1/\left(  N-k\right)  <1$.
\end{The}

\textbf{Proof: }To discuss the stability of the level-independent QBD process
$Q$, we will apply the mean drift method given Neuts \cite{Neu:1981} or Li
\cite{Li:2010}. To this end, we write%
\begin{align*}
\mathbf{A}  &  =A_{0}+A_{1}+A_{2}\\
&  ={\small {=}}\left(
\begin{array}
[c]{ccccc}%
-N\mu & N\mu &  &  & \\
& -\left(  N-1\right)  \mu & \left(  N-1\right)  \mu &  & \\
&  & \ddots & \ddots & \\
&  &  & -(N-2f+1)\mu & (N-2f+1)\mu\\
(N-2f)\mu &  &  &  & -(N-2f)\mu
\end{array}
\right)  .
\end{align*}
It is clear that the Markov chain $\mathbf{A}$ is irreducible, aperiodic and
positive recurrence due to the fact that its state space is finite and
$\mathbf{A}e=0$, where $e$ is a column vector with all components ones. In
this case, let $\mathbf{\theta}=(\theta_{0},\theta_{1},\ldots,\theta
_{2f-1},\theta_{2f})$ be the stationary probability vector of the Markov chain
$\mathbf{A}$. Therefore, the stationary probability vector $\theta$ satisfies
the system of linear equations: $\theta A=0$ and $\theta e=1$. Based on this,
we obtain%
\begin{align*}
-N\theta_{0}+\left(  N-2f\right)  \theta_{2f}  &  =0,\\
\left(  N-k\right)  \theta_{k}-\left(  N-k-1\right)  \theta_{k+1}  &
=0,\text{ }1\leq k\leq2f-1.
\end{align*}
Hence, we get%
\[
\theta_{0}=\frac{1}{\sum\limits_{k=0}^{2f}\frac{N}{N-k}}%
\]
and for $1\leq k\leq2f$%
\[
\theta_{k}=\frac{N}{N-k}\theta_{0}=\frac{N}{N-k}\frac{1}{\sum\limits_{k=0}%
^{2f}\frac{N}{N-k}}.
\]

By using the mean drift method, it is easy to see that the level-independent
QBD process $Q$ is positive recurrent if and only if%

\[
\mathbf{\theta}A_{0}e<\mathbf{\theta}A_{2}e.
\]
It is easy to check that%
\[
\mathbf{\theta}A_{0}e=\lambda\mathbf{\theta}Ie=\lambda,
\]
where $I$ is an identity matrix. At the same time, we have%
\[
\mathbf{\theta}A_{2}e=\theta_{2f}\left(  N-2f\right)  \mu=\frac{\mu}%
{\sum\limits_{k=0}^{2f}\frac{1}{N-k}}.
\]
This gives%
\[
\frac{\mu}{\sum\limits_{k=0}^{2f}\frac{1}{N-k}}<\lambda,
\]
that is,%
\[
\rho=\frac{\lambda}{\mu}\sum_{k=0}^{2f}\frac{1}{N-k}<1.
\]
This completes the proof. \textbf{{\rule{0.08in}{0.08in}}}

In what follows we compute the stationary probability vector of the QBD
process $Q$. Based on this, we can set up four useful performance measures of
the PBFT consensus mechanism.

If $\rho=\left(  \lambda/\mu\right)  \sum_{k=0}^{2f}1/\left(  N-k\right)  <1$,
then the QBD process $Q$ is irreducible and positive recurrent. In this case,
Let $\pi=(\pi_{0},\pi_{1},\pi_{2},\pi_{3},\ldots)$ be the stationary
probability vector of the QBD process $Q$, where%
\[
\pi_{0}=(\pi_{0,0})
\]
and for $k=1,2,3,\ldots$,%
\[
\pi_{k}=(\pi_{k,0},\pi_{k,1},\ldots,\pi_{k,2f-1},\pi_{k,2f}).
\]
Then the vector $\pi$ uniquely satisfies the system of linear equations: $\pi
Q=0$ and $\pi e=1$. Note that the QBD process $Q$ is level-independent, thus
it follows from Chapter 3 of Neuts \cite{Neu:1981} that the stationary
probability vector $\pi$ is a matrix-geometric solution.

The following theorem is a direct result of the level-independent QBD process
$Q$ by using Chapter 3 of Neuts \cite{Neu:1981}, thus we restate it but its
proof is omitted here.

\begin{The}
If $\rho=\left(  \lambda/\mu\right)  \sum_{k=0}^{2f}1/\left(  N-k\right)  <1$,
then the QBD process $Q$ is irreducible and positive recurrent, and its
stationary probability vector $\pi=(\pi_{0},\pi_{1},\pi_{2},\pi_{3},\ldots)$
is given by%
\[
\pi_{k}=\pi_{1}R^{k-1},\text{ }k\geq1,
\]
where $\pi_{0}$ and $\pi_{1}$ can be uniquely determined by the following
system of linear equations%
\[
\pi_{0}B_{1}+\pi_{1}B_{2}=0,
\]%
\[
\pi_{0}B_{0}+\pi_{1}(A_{1}+RA_{2})=0,
\]%
\[
\pi_{0}+\pi_{1}(I-R)^{-1}e=1.
\]
\end{The}

Note that the stationary probability vector $\pi=(\pi_{0},\pi_{1},\pi_{2}%
,\pi_{3},\ldots)$ in general has not an explicit expression, thus we mainly
develop the numerical solution to the vector $\pi$. To this end, it is easy to
see from Chapter 3 of Neuts \cite{Neu:1981} that we first need to numerically
compute the rate matrix $R$, which is the minimal nonnegative solution to the
following nonlinear matrix eqution%

\[
R^{2}A_{2}+RA_{1}+A_{0}=0.
\]
Based on this, the rate matrix $R$ can be approximately calculated by an
iterative algorithm as follows:%

\[
R_{0}=0,
\]
and for $n=0,1,2,\ldots,$%

\[
R_{n+1}=(R_{n}^{2}A_{2}+A_{0})(-A_{1})^{-1}.
\]
For the matrix sequence $\left\{  R\left(  n\right)  ;n\geq0\right\}  $,
Chapter 3 of Neuts \cite{Neu:1981} indicated that $R\left(  n\right)  \uparrow
R$ as $n\rightarrow\infty$. For any sufficiently small positive number
$\varepsilon$ within the desired degree of accuracy set at $10^{-12}$, if
there exists a positive integer $n$ such that
\[
\left\|  R\left(  n+1\right)  -R\left(  n\right)  \right\|  =\max_{0\leq
i\leq2f}\left\{  \sum\limits_{j=0}^{2f}\left|  R\left(  n+1\right)
_{i,j}-R\left(  n\right)  _{i,j}\right|  \right\}  <\varepsilon,
\]
then we take $R=R\left(  n\right)  $. This gives an approximate solution of
the stationary probability vector $\pi$ by means of Theorem 2.

Once the stationary probability vector $\pi$ is computed numerically, we can
provide four useful performance measures of the PBFT consensus mechanism as follows:

\textbf{(a) The average stationary number of transaction packages at the client}%

\[
E\left[  K\right]  =\underset{k=1}{\overset{\infty}{\sum}}k\pi_{k}%
e=\underset{k=1}{\overset{\infty}{\sum}}k\pi_{1}R^{k-1}e=\pi_{1}(I-R)^{-2}e.
\]

\textbf{(b) The average stationary number of nodes to have verified the
validity of a transaction package}%

\[
E\left[  M\right]  =\underset{k=1}{\overset{\infty}{\sum}}\underset
{m=0}{\overset{2f}{\sum}}m\pi_{k,m}=\underset{k=1}{\overset{\infty}{\sum}}%
\pi_{k}\mathbf{\phi}=\pi_{1}(I-R)^{-1}\mathbf{\phi},
\]
where $\mathbf{\phi}=(0,1,2,\ldots,2f-1,2f)^{T}$.

\textbf{(c) The stationary block-pegged rate of the PBFT consensus mechanism}

From Figure 1, it is seen that the block-generated and block-pegged processes
are described as a queueing process, in which the inputs are a Poisson process
but the service process is relatively complicated.

From Figure 2, we can observe that the block-pegged process is a Markovian
arrival process (MAP) whose matrix representation $(C,D)$ of infinite sizes
are given by%
\[
C=\left(
\begin{array}
[c]{cccccc}%
B_{1} & B_{0} &  &  &  & \\
& A_{1} & A_{0} &  &  & \\
&  & A_{1} & A_{0} &  & \\
&  &  & A_{1} & A_{0} & \\
&  &  &  & \ddots & \ddots
\end{array}
\right)  ,\text{ }D=\left(
\begin{array}
[c]{ccccc}%
0 &  &  &  & \\
B_{2} & 0 &  &  & \\
& A_{2} & 0 &  & \\
&  & A_{2} & 0 & \\
&  &  & \ddots & \ddots
\end{array}
\right)  .
\]
It is clear that $Q=C+D$.

If $\rho=\left(  \lambda/\mu\right)  \sum_{k=0}^{2f}1/\left(  N-k\right)  <1$,
then the QBD process $Q$ is irreducible and positive recurrent, and the
stationary block-pegged rate of the PBFT consensus mechanism is given by
\begin{align*}
\gamma &  =\pi De=\pi_{1}B_{2}e+\left(  \sum\limits_{k=2}^{\infty}\pi
_{k}\right)  A_{2}e\\
&  =\pi_{1}B_{2}e+\pi_{1}R\left(  I-R\right)  ^{-1}A_{2}e.
\end{align*}

\textbf{(d) The stationary block-pegged reward per unit time of the major node}

Note that the $N$ nodes are equal as a major node in the PBFT consensus
mechanism, thus any one of them becomes the major node with probability $1/N$.
It is easy to see that the stationary block-pegged total reward per unit time
of the PBFT consensus mechanism is $\gamma c$, thus the stationary
block-pegged reward per unit time of the major node is given by%
\[
\Upsilon=\frac{1}{N}\gamma c=\frac{1}{N}\left[  \pi_{1}B_{2}e+\pi_{1}R\left(
I-R\right)  ^{-1}A_{2}e\right]  c.
\]

\section{Numerical Analysis}

In this section, we use two groups of numerical examples to verify the
validity of our theoretical results, and to show how the four performance
measures of the PBFT consensus mechanism depend on some key system parameters.

\textbf{Group one: }The number $f$ of Byzantine nodes, and the rate $\mu$ of
going through these three stages (\textit{prepare}, \textit{commit}
and\ \textit{reply}).

The system parameters are taken as follows: $\lambda=1$; $\mu\in(3,9)$;
$c=12.5$BTC; $f=50,100,320$; and using $N=3f+1$ leads to $N=151,301,961$.

\begin{figure}[th]
\setlength{\abovecaptionskip}{0.cm}  \setlength{\belowcaptionskip}{-0.cm}
\centering              \includegraphics[width=7cm]{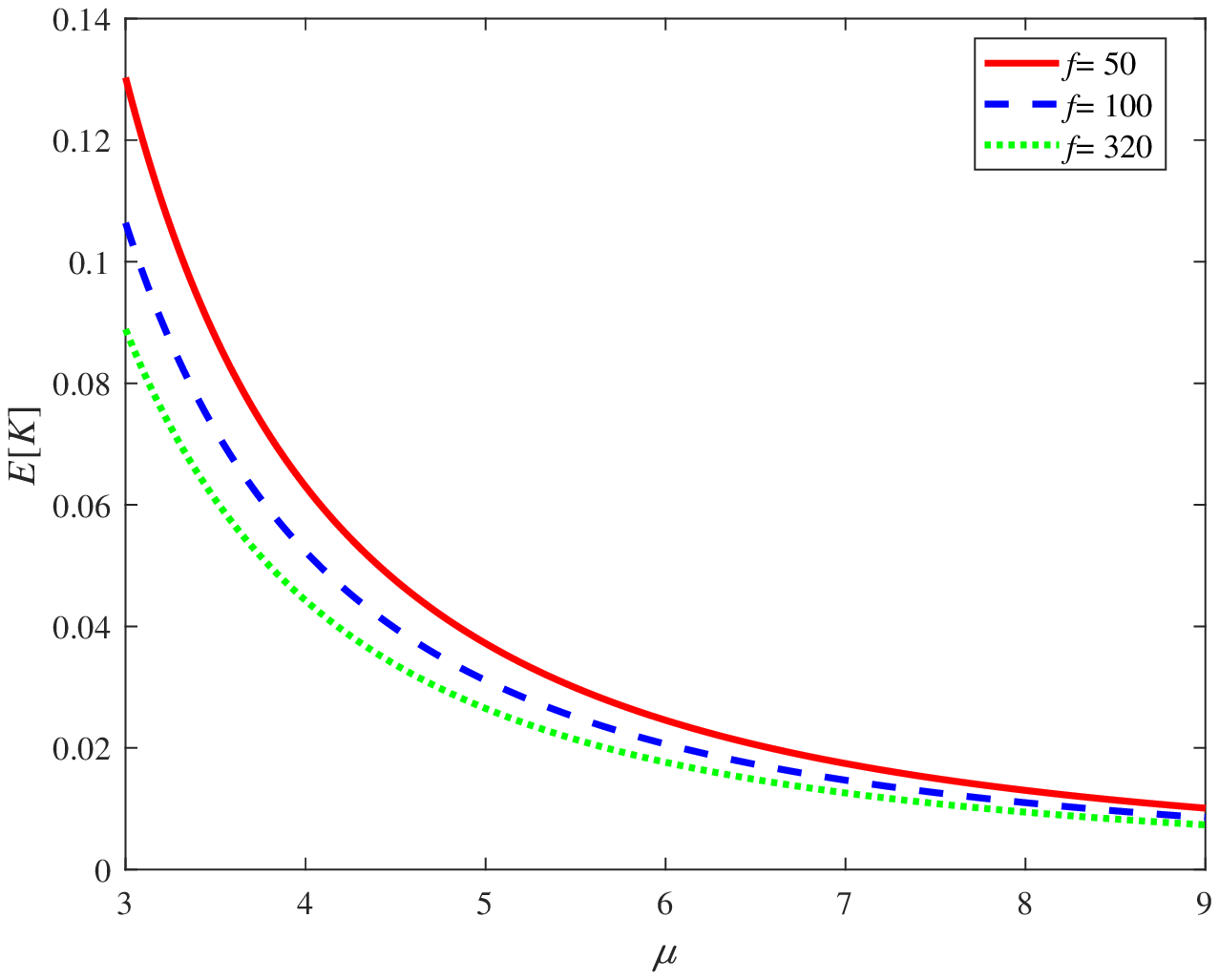}
\includegraphics[width=7cm]{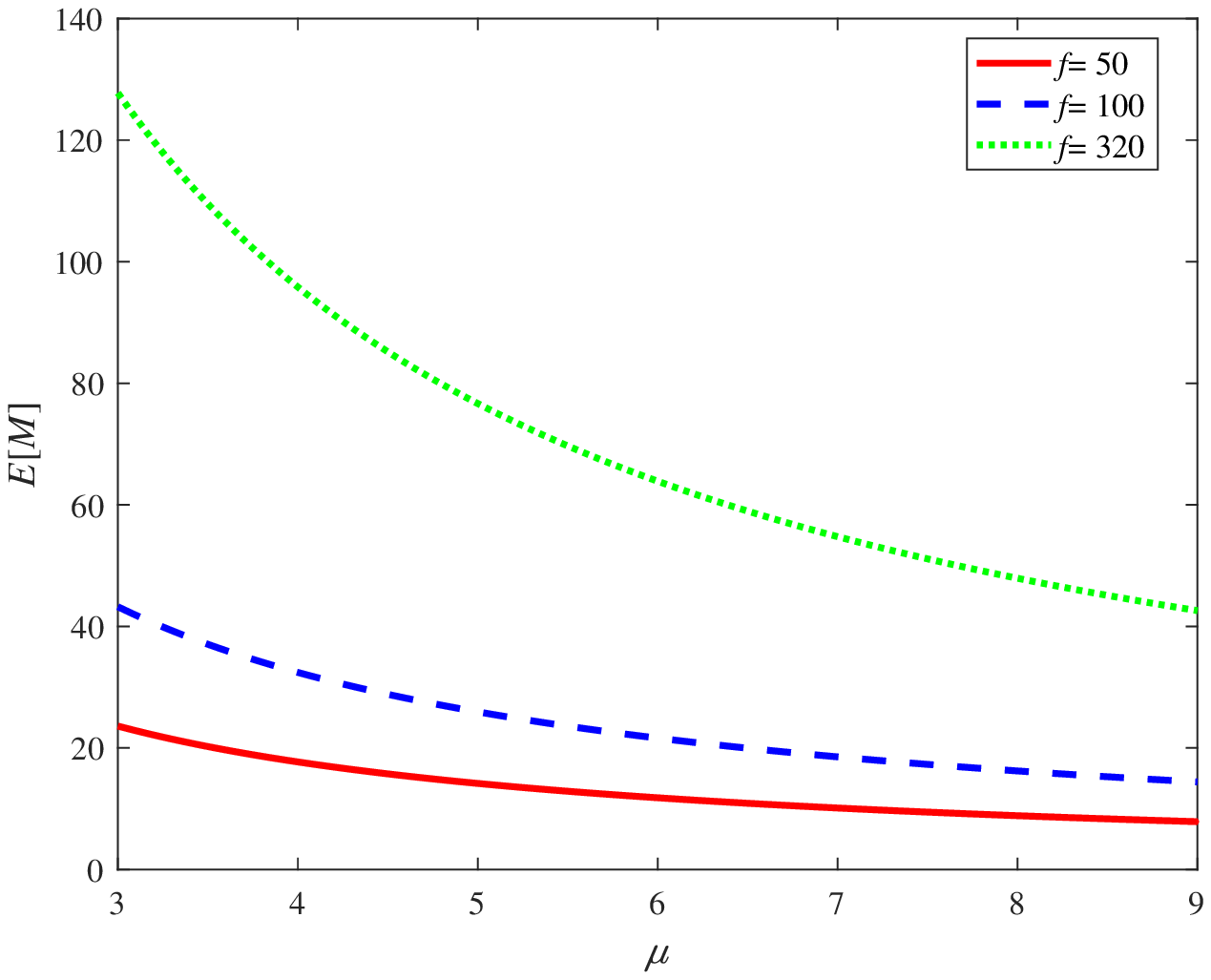}  \newline \caption{$E[K]$ and $E[M]$
vs. $\mu$ and $f$}%
\label{figure:figure-3}%
\end{figure}

From the left of Figure 3, it is seen that $E[K]$ decreases as $\mu$
increases, and it also decreases as $f$ (or $N$) increases. This shows that
the PBFT consensus mechanism makes the block-pegged on blockchain faster
either as $\mu$ increases or as $f$ (or $N$) increases. Such a numerical
result as $\mu$ increases is intuitive from a real observation; while another
result as $f$ (or $N$) increases has an interesting practical significance,
i.e., we can speed up the block-pegged on blockchain through increasing the
number $N$ of nodes in the blockchain network. Similarly, it is observed from
the right of Figure 3 that $E[M]$ decreases as $\mu$ increases; but it
increases as $f$ (or $N$) increases, this can easily be understood due to the
increase of $N$.

\begin{figure}[th]
\setlength{\abovecaptionskip}{0.cm}  \setlength{\belowcaptionskip}{-0.cm}
\centering               \includegraphics[width=7cm]{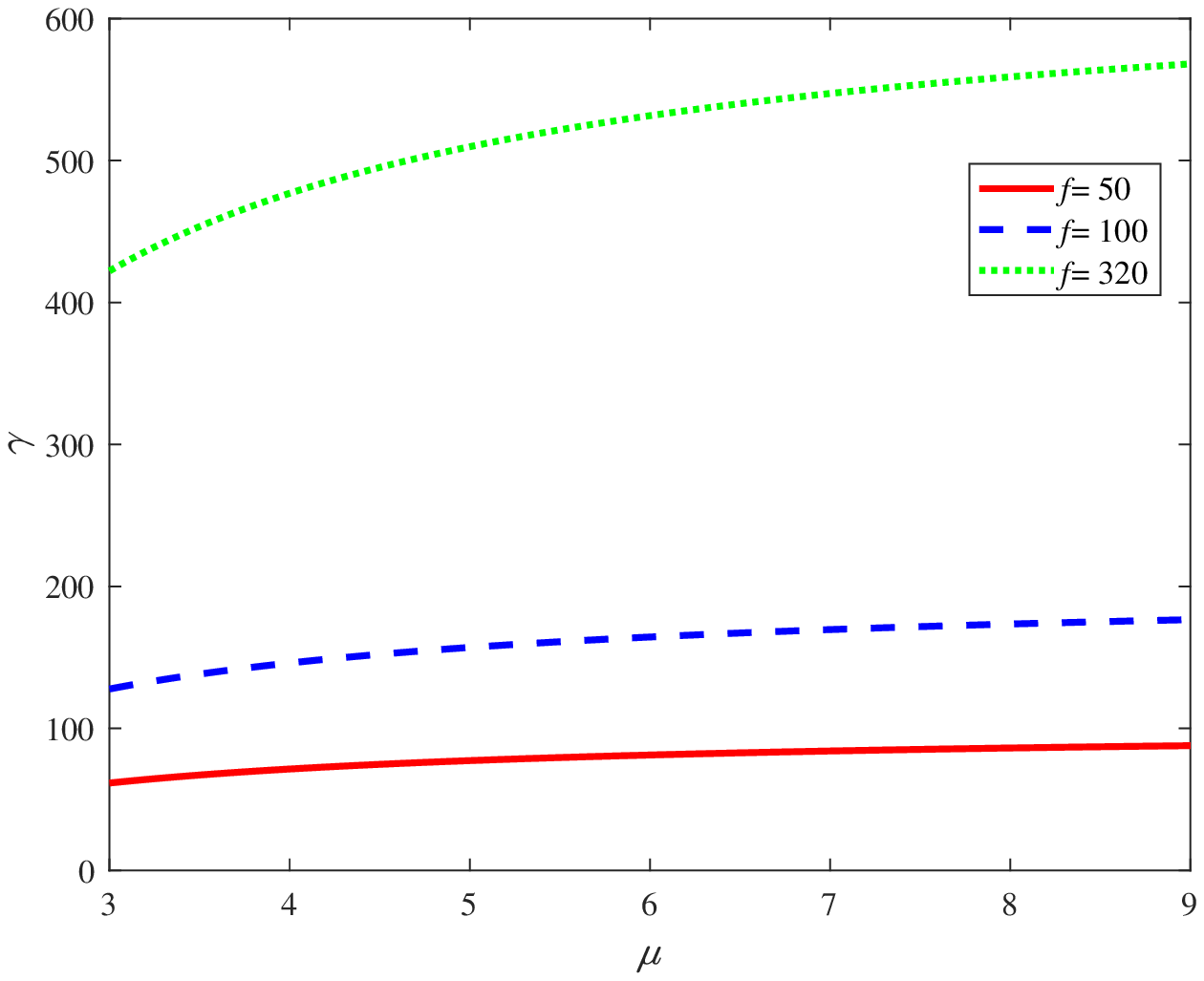}
\includegraphics[width=7cm]{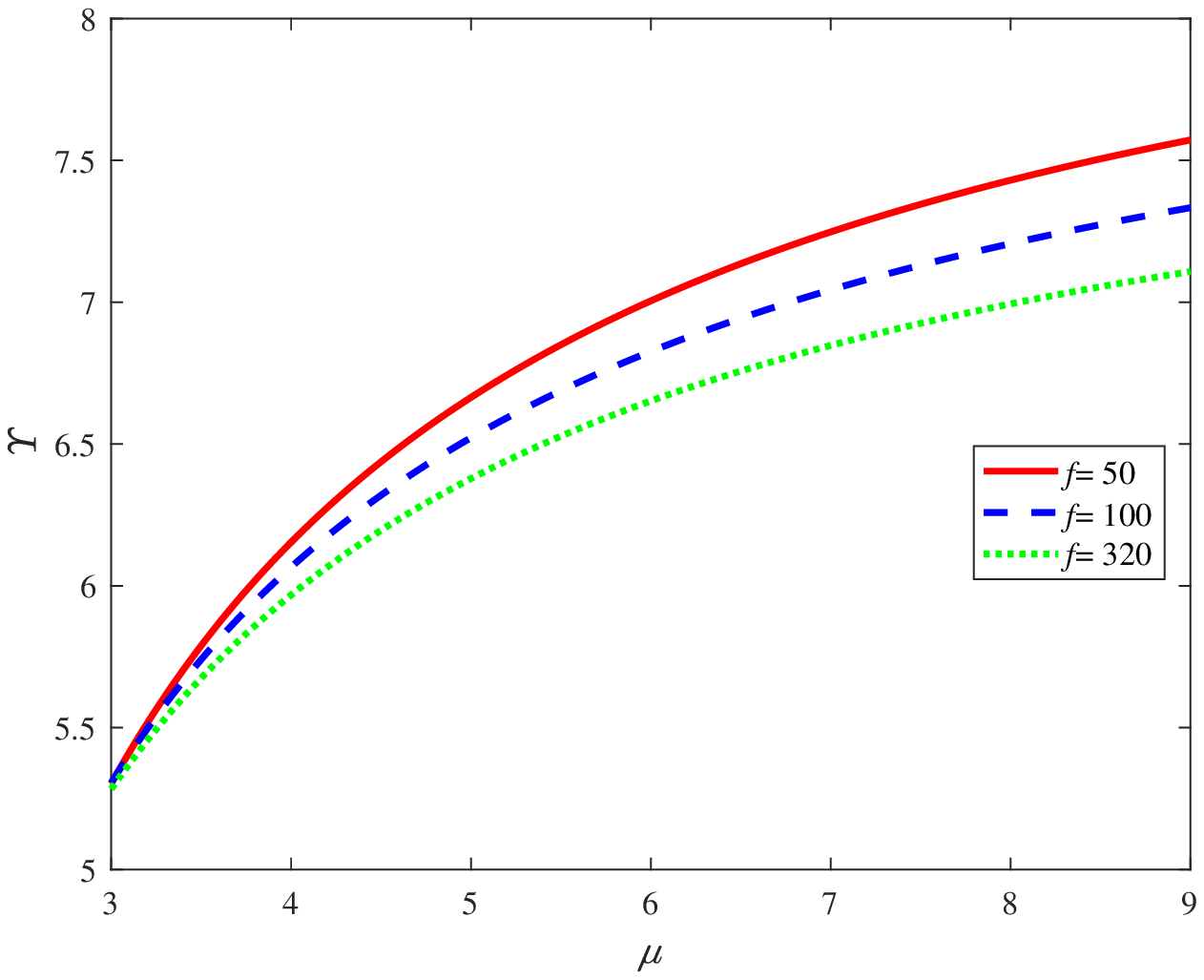}  \newline \caption{$\gamma$ and
$\Upsilon$ vs. $\mu$ and $f$}%
\label{figure:figure-4}%
\end{figure}

From the left of Figure 4, it is seen that $\gamma$ increases as $\mu$
increases, and it also increases as $f$ (or $N$) increases. This shows that
the PBFT consensus mechanism makes the block-pegged on blockchain faster
either as $\mu$ increases or as $f$ (or $N$) increases. Such two numerical
results are intuitive. Similarly, it is observed from the right of Figure 4
that $\Upsilon$ increases as $\mu$ increases; but it decreases as $f$ (or $N$)
increases, this shows that a bigger number $N$ can increase the profit of each node.

\textbf{Group two: }The number $f$ of Byzantine nodes, and the arrival rate
$\lambda$ of transaction packages at the client.

The system parameters are taken as follows: $\mu=9$; $\lambda\in(1,3)$;
$c=12.5$BTC; $f=50,100,320$; and using $N=3f+1$ leads to $N=151,301,961$.

\begin{figure}[th]
\setlength{\abovecaptionskip}{0.cm}  \setlength{\belowcaptionskip}{-0.cm}
\centering              \includegraphics[width=7cm]{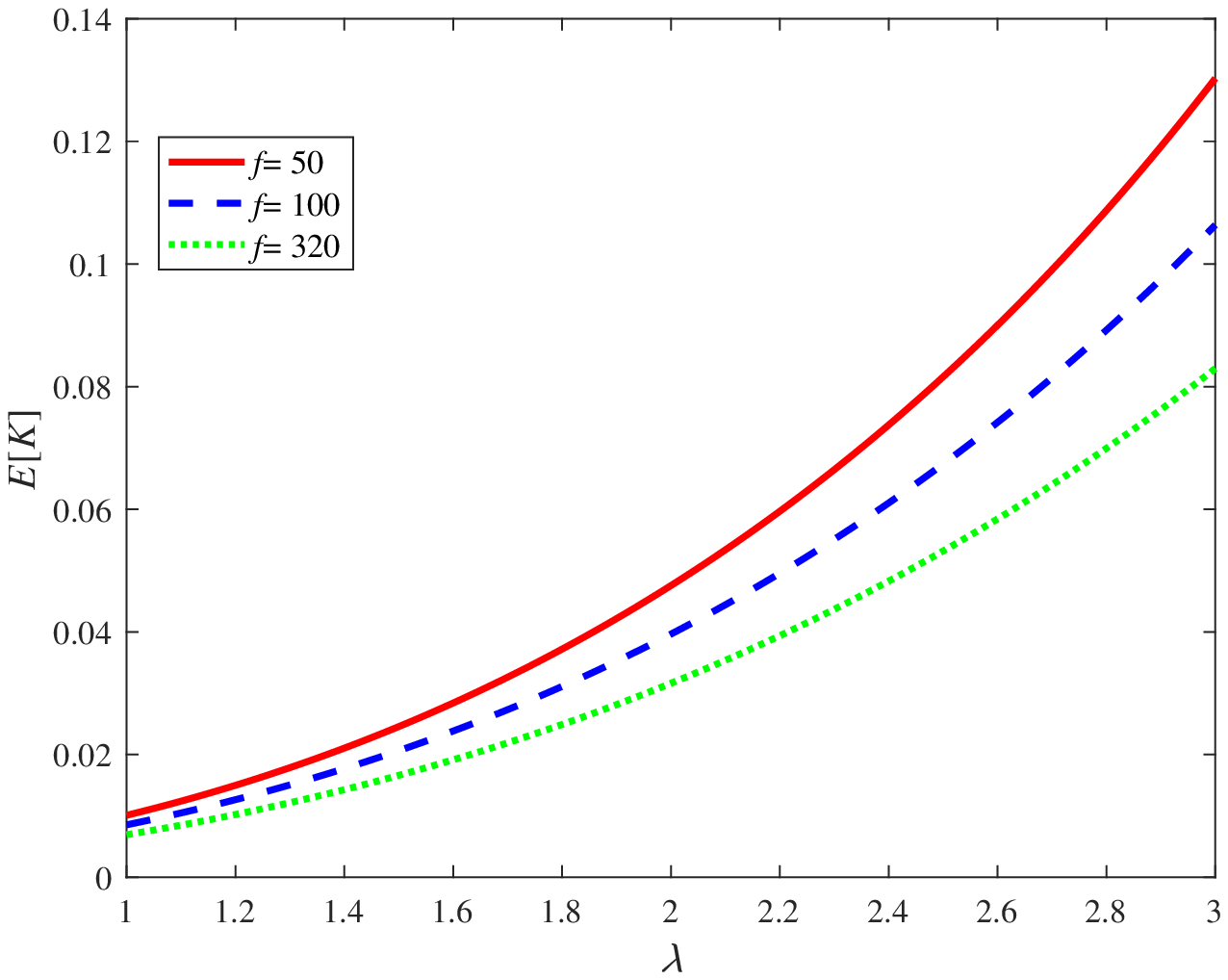}
\includegraphics[width=7cm]{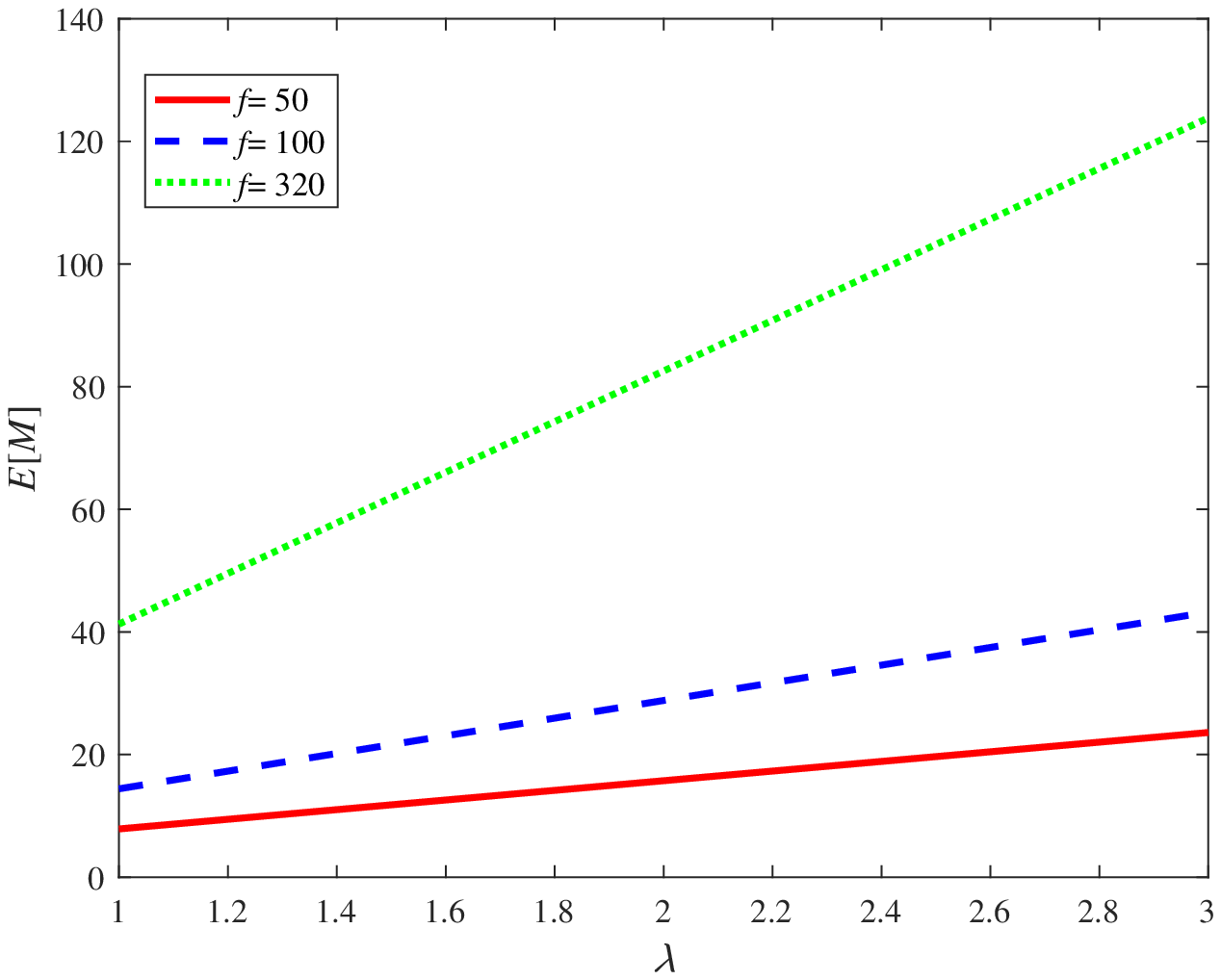}  \newline \caption{$E[K]$ and $E[M]$
vs. $\lambda$ and $f$}%
\label{figure:figure-5}%
\end{figure}

From the left of Figure 5, it is seen that $E[K]$ increases as $\lambda$
increases, but it decreases as $f$ (or $N$) increases. This shows that the
PBFT consensus mechanism makes the block-pegged on blockchain faster either as
$\lambda$ increases or as $f$ (or $N$) increases. Similarly, it is observed
from the right of Figure 5 that $E[M]$ increases as $\lambda$ increases; and
it also increases as $f$ (or $N$) increases, this can easily be explained by
increasing the number $N$.

\begin{figure}[th]
\setlength{\abovecaptionskip}{0.cm}  \setlength{\belowcaptionskip}{-0.cm}
\centering               \includegraphics[width=7cm]{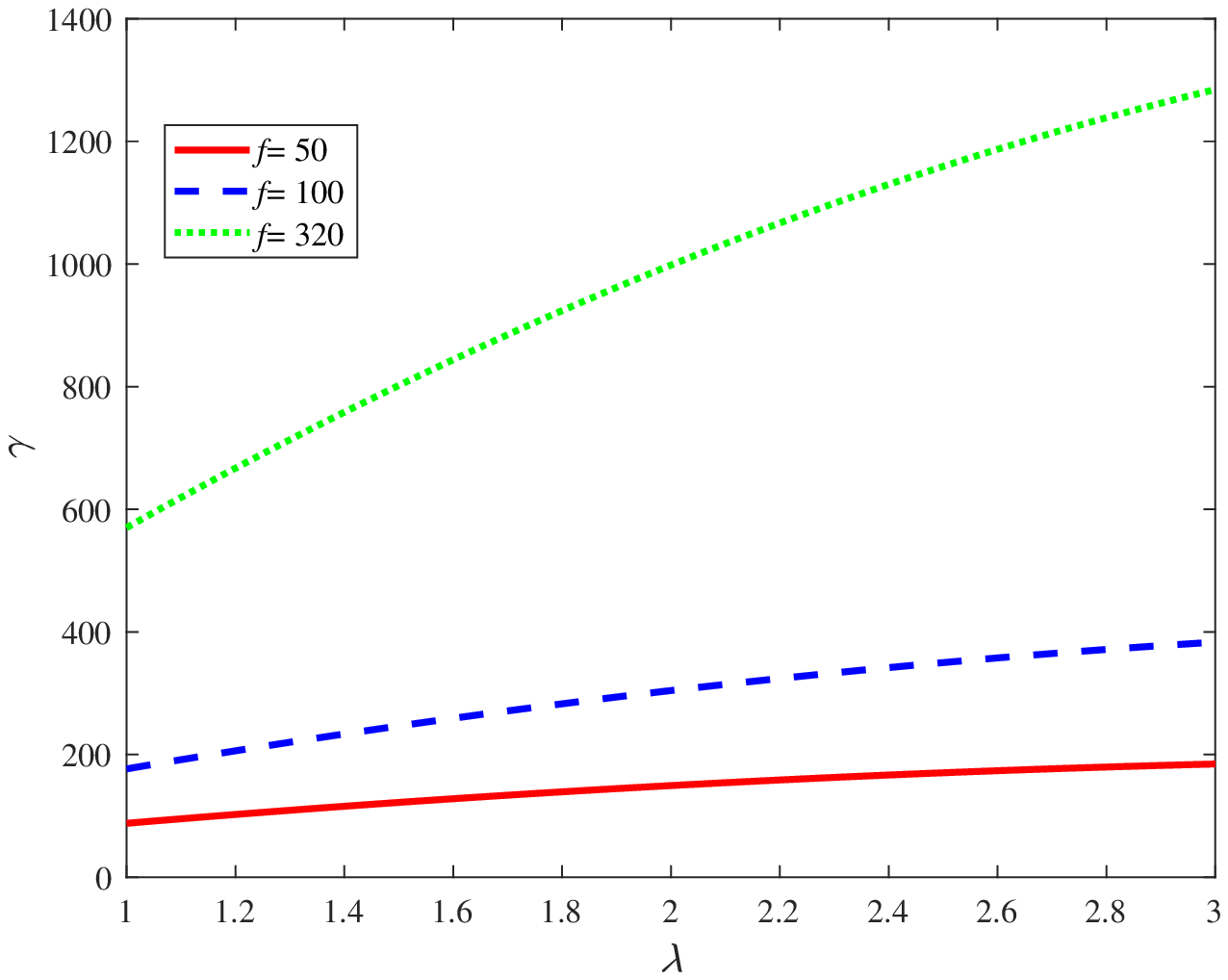}
\includegraphics[width=7cm]{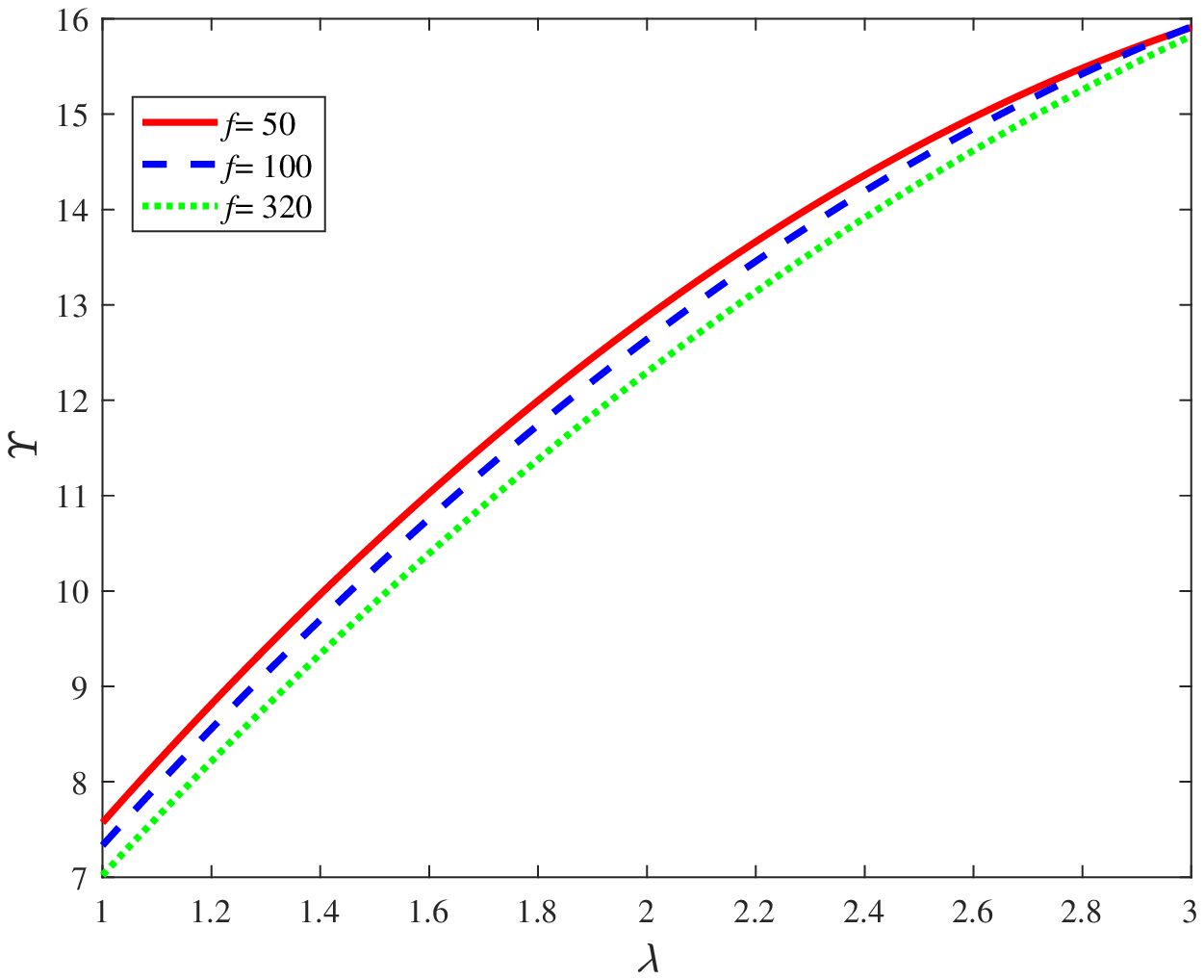}  \newline \caption{$\gamma$ and
$\Upsilon$ vs. $\lambda$ and $f$}%
\label{figure:figure-6}%
\end{figure}

From the left of Figure 6, it is seen that $\gamma$ increases as $\lambda$
increases, and it also increases as $f$ (or $N$) increases. This shows that
the PBFT consensus mechanism makes the block-pegged on blockchain faster
either as $\mu$ increases or as $f$ (or $N$) increases. Similarly, it is
observed from the right of Figure 6 that $\Upsilon$ increases as $\mu$
increases; but it decreases as $f$ (or $N$) increases, this shows that a
bigger number $N$ can increase the profit of each node.

\section{Concluding Remarks}

In this paper, we describe a simple stochastic performance model of the PBFT
consensus mechanism, and refine the stochastic performance model as not only a
queueing system with complicated service times but also a level-independent
QBD process. We apply the matrix-geometric solution to obtain a necessary and
sufficient condition under which the PBFT consensus system is stable, and
establish the stationary probability vector of the QBD process. Based on this,
we provide four useful performance measures of the PBFT consensus mechanism,
and can numerically compute each of them. Finally, we use some numerical
examples to show how the four performance measures are influenced by some key
parameters of the PBFT consensus mechanism. By means of the theory of
multi-dimensional Markov processes, we are optimistic that the methodology and
results given in this paper are applicable in a wide range research of PBFT
consensus mechanism and even other types of consensus mechanisms.

Along this line, we will continue our future research on several interesting
directions as follows:

---- Let all the three stages (\textit{prepare}, \textit{commit}
and\ \textit{reply}) be differently exponential distributions with means
$1/\mu_{1}$, $1/\mu_{2}$ and $1/\mu_{3}$, respectively. Note that the extended
stochastic performance model is far more difficult than that of this paper due
to some complicated phase-type calculation.

--- Developing effective algorithms for more general stochastic performance
model with the Markovian arrival process of transaction packages at the
client, and let all the three stages (\textit{prepare}, \textit{commit}
and\ \textit{reply}) be phase-type (PH) distributions with matrix
representations $\left(  \alpha_{1},T_{1}\right)  $, $\left(  \alpha_{2}%
,T_{2}\right)  $ and $\left(  \alpha_{3},T_{3}\right)  $, respectively.

--- When the arrivals of transaction packages at the client are a renewal
process, and/or some of the three stages (\textit{prepare}, \textit{commit}
and\ \textit{reply}) follow general probability distributions, an interesting
future research is to focus on fluid and diffusion approximations of PBFT
consensus mechanism.

--- Setting up reward function with respect to cost structure, transaction
fee, mining reward, security and so forth. It is very interesting in our
future study to develop stochastic optimization, Markov decision processes and
stochastic game models in the study of PBFT consensus mechanism and even other
types of consensus mechanisms.

\section*{Acknowledgment}

Quan-Lin Li was supported by the National Natural Science Foundation of China
under grants No. 71671158 and 71932002.


\vskip     3cm

\begin{thebibliography}{9}                                                                                                %

\bibitem {Abr:2017}Abraham, I., Gueta, G., Malkhi, D., Alvisi, L., Kotla, R.,
\& Martin, J. P. (2017). Revisiting fast practical byzantine fault tolerance.
arXiv preprint arXiv:1712.01367.

\bibitem {Alq:2021}Alqahtani, S., \& Demirbas, M. (2021). Bottlenecks in
blockchain consensus protocols. arXiv preprint arXiv:2103.04234.

\bibitem {Bai:2019}Bai, Q., Zhou, X., Wang, X., Xu, Y., Wang, X., \& Kong, Q.
(2019). A deep dive into blockchain selfish mining. In: \textit{The 2019 IEEE
International Conference on Communications}, pp. 1-6.

\bibitem {Ban:2017}Bano, S., Sonnino, A., Al-Bassam, M., Azouvi, S., McCorry,
P., Meiklejohn, S., \& Danezis, G. (2017). Consensus in the age of
blockchains. arXiv preprint arXiv:1711.03936.

\bibitem {Bas:2018}Bashir, I. (2018). \textit{Mastering Blockchain:
Distributed Ledger Technology, Decentralization, and Smart Contracts
Explained}. Packt Publishing Ltd.

\bibitem {Bra:2020}Bravo, M., Istv\'{a}n, Z., \& Sit, M. K. (2020). Towards
improving the performance of BFT consensus for future permissioned
blockchains. arXiv preprint arXiv:2007.12637.

\bibitem {Cac:2017}Cachin, C., \& Vukoli\'{c}, M. (2017). Blockchain consensus
protocols in the wild. arXiv preprint arXiv:1707.01873.

\bibitem {Carlsten:2016}Carlsten, M. (2016). The impact of transaction fees on
bitcoin mining strategies. Doctoral Dissertation, Princeton University.

\bibitem {Car:2020}Carrara, G. R., Burle, L. M., Medeiros, D. S., de
Albuquerque, C. V. N., \& Mattos, D. M. (2020). Consistency, availability, and
partition tolerance in blockchain: a survey on the consensus mechanism over
peer-to-peer networking. \textit{Annals of Telecommunications}, 75(3), 163-174.

\bibitem {Castro:1999}Castro, M., \& Liskov, B. (1999). Practical Byzantine
fault tolerance. In: \textit{Proceedings of the Third Symposium on Operating
Systems Design and Implementation}, pp: 173-186.

\bibitem {Cas:2002}Castro, M., \& Liskov, B. (2002). Practical Byzantine fault
tolerance and proactive recovery. \textit{ACM Transactions on Computer
Systems}, 20(4), 398-461.

\bibitem {Cha:2018}Chaudhry, N., \& Yousaf, M. M. (2018). Consensus algorithms
in blockchain: Comparative analysis, challenges and opportunities. In:
\textit{The 12th International Conference on Open Source Systems and
Technologies}, pp. 54-63.

\bibitem {Dai:2020}Dai, H. N., Zheng, Z., \& Zhang, Y. (2019). Blockchain for
Internet of Things: A survey. \textit{IEEE Internet of Things Journal}, 6(5), 8076-8094.

\bibitem {Ekr:2020}Ekramifard, A., Amintoosi, H., Seno, A. H., Dehghantanha,
A., \& Parizi, R. M. (2020). A systematic literature review of integration of
blockchain and artificial intelligence. In: \textit{Blockchain Cybersecurity,
Trust and Privacy}, pp. 147-160. Springer, Advances in Information Security
book series, volume 79.

\bibitem {Fan:2020}Fan, C., Ghaemi, S., Khazaei, H., \& Musilek, P. (2020).
Performance evaluation of blockchain systems: A systematic survey.
\textit{IEEE Access}, 8, 126927-126950.

\bibitem {FanM:2020}Fang, M., \& Liu, J. (2020). Toward low-cost and stable
blockchain networks. In: \textit{The 2020 IEEE International Conference on
Communications}, pp. 1-6.

\bibitem {Fau:2020}Fauziah, Z., Latifah, H., Omar, X., Khoirunisa, A., \&
Millah, S. (2020). Application of blockchain technology in smart contracts: A
systematic literature review. \textit{Aptisi Transactions on
Technopreneurship}, 2(2), 160-166.

\bibitem {Fer:2020}Ferdous, M. S., Chowdhury, M. J. M., Hoque, M. A., \&
Colman, A. (2020). Blockchain consensuses algorithms: A survey. arXiv preprint arXiv:2001.07091.

\bibitem {Fra:2020}Fralix, B. (2020). On classes of Bitcoin-inspired
infinite-server queueing systems. \textit{Queueing Systems}, 95, 29--52.

\bibitem {Gei:2019}Geissler, S., Prantl, T., Lange, S., Wamser, F., \&
Hossfeld, T. (2019). Discrete-time analysis of the blockchain distributed
ledger technology. In: \textit{The 31st International Teletraffic Congress},
pp. 130-137.

\bibitem {Goebel:2016}G\"{o}bel, J., Keeler, H. P., Krzesinski, A. E., \&
Taylor, P. G. (2016). Bitcoin blockchain dynamics: The selfish-mine strategy
in the presence of propagation delay. \textit{Performance Evaluation}, 104, 23-41.

\bibitem {Gop:2020}Gopalan, A., Sankararaman, A., Walid, A., \& Vishwanath, S.
(2020). Stability and scalability of blockchain systems. \textit{Proceedings
of the ACM on Measurement and Analysis of Computing Systems}, 4(2), 1-35.

\bibitem {Gueta:2019}Gueta, G. G., Abraham, I., Grossman, S., et al. (2019).
Sbft: a scalable and decentralized trust infrastructure. In: \textit{The 49th
Annual IEEE/IFIP International Conference on Dependable Systems and Networks},
pp. 568-580.

\bibitem {Hao:2018}Hao, X., Yu, L., Liu, Z., Zhen, L., \& Dawu, G. (2018).
Dynamic practical byzantine fault tolerance. In: \textit{The 2018 IEEE
Conference on Communications and Network Security}, pp. 1-8.

\bibitem {Huang:2019}Huang, D., Ma, X., \& Zhang, S. (2019). Performance
analysis of the raft consensus algorithm for private blockchains. \textit{IEEE
Transactions on Systems, Man, and Cybernetics: Systems}, 50(1), 172-181.

\bibitem {Hua:2021}Huang, H., Kong, W., Zhou, S., Zheng, Z., \& Guo, S.
(2021). A survey of state-of-the-art on blockchains: Theories, modelings, and
tools. \textit{ACM Computing Surveys}, 54(2), 1-42.

\bibitem {Jav:2020}Javier, K. and Fralix, B. (2020). A further study of some
Markovian Bitcoin models from G\"{o}bel et al.. \textit{Stochastic Models},
36(2), 223-250.

\bibitem {Kha:2021}Khamar, J., \& Patel, H. (2021). An extensive survey on
consensus mechanisms for blockchain technology. In: \textit{Data Science and
Intelligent Applications}, pp. 363-374. Springer, Lecture Notes on Data
Engineering and Communications Technologies book series, volume 52.

\bibitem {Kiayias:2018}Kiayias, A., \& Russell, A. (2018). Ouroboros-BFT: A
simple Byzantine fault tolerant consensus protocol. IACR Cryptol. ePrint Arch.
1049. https://eprint.iacr.org/2018/1049.pdf.

\bibitem {Kiffer:2018}Kiffer, L., Rajaraman, R., \& Shelat, A. (2018). A
better method to analyze blockchain consistency. In: \textit{Proceedings of
the 2018 ACM SIGSAC Conference on Computer and Communications Security}, pp. 729-744.

\bibitem {Lam:1983}Lamport, L. (1983). The weak Byzantine generals problem.
\textit{Journal of the ACM}, 30(3), 668-676.

\bibitem {Lam:1982}Lamport L. Shostak R, Pease M. (1982) The Byzantine
generals problem. \textit{ACM Transactions on Programming Languages and
Systems}, 4(3), 382-401.

\bibitem {Li:2010}Li, Q. L. (2010). Constructive Computation in Stochastic
Models with Applications: The RG-Factorizations. Springer.

\bibitem {Li:2020}Li, Q. L., Chang, Y. X., Wu, X., \& Zhang, G. (2020). A new
theoretical framework of pyramid markov processes for blockchain selfish
mining. arXiv preprint arXiv:2007.01459.

\bibitem {Li:2018}Li, Q. L., Ma, J. Y., \& Chang, Y. X. (2018). Blockchain
queue theory. In: \textit{International Conference on Computational Social
Networks}, pp. 25-40. Springer, Lecture Notes in Computer Science book series,
volume 11280.

\bibitem {Li:2019}Li, Q. L., Ma, J. Y., Chang, Y. X., Ma, F. Q., \& Yu, H. B.
(2019). Markov processes in blockchain systems. \textit{Computational Social
Networks}, 6(1), 1-28.

\bibitem {Li:2021}Li, Y., Cao, B., Liang, L., Mao, D., \& Zhang, L. (2021).
Block access control in wireless blockchain network: Design, modeling and
analysis. \textit{IEEE Transactions on Vehicular Technology}, Date of
Publication: 14 June 2021.

\bibitem {LiY:2020}Li, Y., Cao, B., Peng, M., Zhang, L., Zhang, L., Feng, D.,
\& Yu, J. (2020). Direct acyclic graph-based ledger for Internet of Things:
Performance and security analysis. \textit{IEEE/ACM Transactions on
Networking}, 28(4), 1643-1656.

\bibitem {Mal:2020}Maleh, Y., Shojafar, M., Alazab, M., \& Romdhani, I.
(Eds.). (2020). \textit{Blockchain for Cybersecurity and Privacy:
Architectures, Challenges, and Applications}. CRC Press.

\bibitem {Malkhi:2019}Malkhi, D., Nayak, K., \& Ren, L. (2019). Flexible
byzantine fault tolerance. In: \textit{Proceedings of the 2019 ACM SIGSAC
Conference on Computer and Communications Security}, pp. 1041-1053.

\bibitem {Mar:2016}Martin, J. P., \& Alvisi, L. (2006). Fast Byzantine
consensus. \textit{IEEE Transactions on Dependable and Secure Computing},
3(3), 202-215.

\bibitem {Mes:2021}Meshcheryakov, Y., Melman, A., Evsutin, O., Morozov, V., \&
Koucheryavy, Y. (2021). On performance of PBFT for IoT-applications with
constrained devices. arXiv preprint arXiv:2104.05026.

\bibitem {Mis:2020}Mi\v{s}i\'{c}, J., Mi\v{s}i\'{c}, V. B., \& Chang, X.
(2020). Performance of Bitcoin network with synchronizing nodes and a mix of
regular and compact blocks. \textit{IEEE Transactions on Network Science and
Engineering}, 7(4), 3135-3147.

\bibitem {Nar:2016}Narayanan, A., Bonneau, J., Felten, E., Miller, A., \&
Goldfeder, S. (2016). \textit{Bitcoin and Cryptocurrency Technologies: A
Comprehensive Introduction}. Princeton University Press.

\bibitem {Nat:2017}Natoli, C., Yu, J., Gramoli, V., \& Esteves-Verissimo, P.
(2019). Deconstructing blockchains: A comprehensive survey on consensus,
membership and structure. arXiv preprint arXiv:1908.08316.

\bibitem {Nayak:2016}Nayak, K., Kumar, S., Miller, A., \& Shi, E. (2016).
Stubborn mining: Generalizing selfish mining and combining with an eclipse
attack. In: \textit{The 2016 IEEE European Symposium on Security and Privacy},
pp. 305-320.

\bibitem {Neu:1981}Neuts, M. F. (1981). Matrix-Geometric Solutions in
Stochastic Models: An Algorithmic Approach. The Johns Hopkins University Press.

\bibitem {Ngu:2018}Nguyen, G. T., \& Kim, K. (2018). A survey about consensus
algorithms used in blockchain. \textit{Journal of Information processing
systems}, 14(1), 101-128.

\bibitem {Nij:2020}Nijsse, J., \& Litchfield, A. (2020). A taxonomy of
blockchain consensus methods. \textit{Cryptography}, 4(4), 32.

\bibitem {Ongaro:2014}Ongaro, D., \& Ousterhout, J. (2014). In search of an
understandable consensus algorithm. In: \textit{Proceedings of 2014 USENIX
Annual Technical Conference}, pp. 305-319.

\bibitem {Pah:2019}Pahlajani, S., Kshirsagar, A., \& Pachghare, V. (2019).
Survey on private blockchain consensus algorithms. In: \textit{The 1st
International Conference on Innovations in Information and Communication
Technology}, pp. 1-6.

\bibitem {Pap:2018}Papadis, N., Borst, S., Walid, A., Grissa, M., \&
Tassiulas, L. (2018). Stochastic models and wide-area network measurements for
blockchain design and analysis. In: \textit{IEEE INFOCOM 2018-IEEE Conference
on Computer Communications}, pp. 2546-2554.

\bibitem {Pease:1980}Pease, M., Shostak, R., \& Lamport, L. (1980). Reaching
agreement in the presence of faults. \textit{Journal of the ACM}, 27(2), 228-234.

\bibitem {Raj:2019}Raj, K. (2019). \textit{Foundations of Blockchain: The
Pathway to Cryptocurrencies and Decentralized Blockchain Applications}. Packt
Publishing Ltd.

\bibitem {Reh:2020}Rehan, M. M., \& Rehmani, M. H. (Eds.). (2020).
\textit{Blockchain-Enabled Fog and Edge Computing: Concepts, Architectures and
Applications}. CRC Press.

\bibitem {Sak:2020}Sakho, S., Zhang, J., Essaf, F., Badiss, K., Abide, T., \&
Kiprop, J. K. (2020). Research on an improved practical byzantine fault
tolerance algorithm. In: \textit{The 2nd International Conference on Advances
in Computer Technology, Information Science and Communications}, pp. 176-181.

\bibitem {Sal:2018}Salimitari, M., \& Chatterjee, M. (2018). A survey on
consensus protocols in blockchain for iot networks. arXiv preprint arXiv:1809.05613.

\bibitem {Sch:2020}Schar, F., \& Berentsen, A. (2020). \textit{Bitcoin,
Blockchain, and Cryptoassets: A Comprehensive Introduction}. MIT Press.

\bibitem {Schlichting:1983}Schlichting, R. D., \& Schneider, F. B. (1983).
Fail-stop processors: An approach to designing fault-tolerant computing
systems. \textit{ACM Transactions on Computer Systems}, 1(3), 222-238.

\bibitem {Sha:2020}Sharma, P., Jindal, R., \& Borah, M. D. (2020). Blockchain
technology for cloud storage: A systematic literature review. \textit{ACM
Computing Surveys}, 53(4), 1-32.

\bibitem {Sme:2020}Smetanin, S., Ometov, A., Komarov, M., Masek, P., \&
Koucheryavy, Y. (2020). Blockchain evaluation approaches: State-of-the-art and
future perspective. \textit{Sensors}, 20(12), 3358.

\bibitem {Sti:2019}Stifter, N., Judmayer, A., \& Weippl, E. (2019). Revisiting
practical byzantine fault tolerance through blockchain technologies. In:
\textit{Security and Quality in Cyber-Physical Systems Engineering}, pp.
471-495. Springer.

\bibitem {Var:2021}Varma, S. M., \& Maguluri, S. T. (2021). Throughput optimal
routing in blockchain based payment systems. \textit{IEEE Transactions on
Control of Network Systems}, Date of Publication: 14 June 2021.

\bibitem {Veronese:2011}Veronese, G. S., Correia, M., Bessani, A. N., Lung, L.
C., \& Verissimo, P. (2011). Efficient Byzantine fault-tolerance. \textit{IEEE
Transactions on Computers}, 62(1), 16-30.

\bibitem {Wan:2020}Wan, S., Li, M., Liu, G., \& Wang, C. (2020). Recent
advances in consensus protocols for blockchain: a survey. \textit{Wireless
networks}, 26(8), 5579-5593.

\bibitem {Wan:2019}Wang, W., Hoang, D. T., Hu, P. et al. (2019). A survey on
consensus mechanisms and mining strategy management in blockchain networks.
\textit{IEEE Access}, 7, 22328-22370.

\bibitem {Wil:2021}Wilhelmi, F., \& Giupponi, L. (2021). Discrete-time
analysis of wireless blockchain networks. arXiv preprint arXiv:2104.05586.

\bibitem {Xia:2020}Xiao, Y., Zhang, N., Lou, W., \& Hou, Y. T. (2020). A
survey of distributed consensus protocols for blockchain networks.
\textit{IEEE Communications Surveys \& Tutorials}, 22(2), 1432-1465.

\bibitem {Yao:2021}Yao, W., Ye, J., Murimi, R., \& Wang, G. (2021). A survey
on consortium blockchain consensus mechanisms. arXiv preprint arXiv:2102.12058.
\end{thebibliography}
\end{document}